\documentclass[aps,twocolumn,prl,floatfix]{revtex4}
\usepackage{amsmath,amssymb,amsthm,bm,bbm}
\usepackage{graphicx}
\usepackage{siunitx}
\usepackage{braket}
\usepackage{soul}
\usepackage{comment}
\usepackage[normalem]{ulem}
\usepackage{hyperref,url}
\usepackage{overpic}

\usepackage{color}

\newcommand{\myst}[1]{\st{#1}}
\usepackage[normalem]{ulem}

\newcommand{\commutator}[2]{\left[#1,#2\right]}
\renewcommand{\b}[1]{\hat{b}_{#1}}
\newcommand{\bdag}[1]{\hat{b}^{\dagger}_{#1}}
\newcommand{\Hamiltonian}[1]{\hat{H}_{#1}}

\newcommand{\pauli}[2]{\hat{\sigma}_{#1}^{#2}}

\newcommand{\1}{\hat{\mathbbm{1}}}

\newcommand {\be}{\begin{equation}}
\newcommand {\ee}{\end{equation}}
\newcommand {\ba}{\begin{eqnarray}}
\newcommand {\ea}{\end{eqnarray}}

\begin{document}
\title{Quantifying Complexity in Quantum Phase Transitions via Mutual Information Complex Networks}
\author{Marc Andrew Valdez}
\author{Daniel Jaschke}
\author{David L. Vargas}
\author{Lincoln D. Carr}
\affiliation{Department of Physics, Colorado School of Mines, Golden, CO 80401, USA}
\date{\today}

\begin{abstract}
We quantify the emergent complexity of quantum states near quantum critical points on regular 1D lattices, via complex network measures based on quantum mutual information as the adjacency matrix, in direct analogy to quantifying the complexity of EEG/fMRI measurements of the brain.  Using matrix product state methods, we show that network density, clustering, disparity, and Pearson's correlation obtain the critical point for both quantum Ising and Bose-Hubbard models to a high degree of accuracy in finite-size scaling for three classes of quantum phase transitions, $Z_2$, mean field superfluid/Mott insulator, and a BKT crossover.


\end{abstract}

\maketitle

Classical statistical physics has developed a powerful set of tools for analyzing complex systems,
chief among them complex networks, in which connectivity and topology predominate over other
system features~\cite{Newman2003}.  Complex networks model systems as diverse as the brain and
the internet; however, up till now they have been obtained in quantum systems by explicitly enforcing complex network structure in their quantum connections~\cite{Bianconi2001,Kimble2008,Cuquet2009,Perseguers2010,Bianconi2012a,Bianconi2012b}, e.g. entanglement percolation \emph{on} a complex network~\cite{Cuquet2009}.  In contrast, complexity measures on the brain observe emergent complexity arising \emph{out of}, e.g., a regular array of EEG electrodes placed on the scalp, via an adjacency matrix formed from the classical mutual information calculated between them~\cite{Bullmore2009}.  We apply the quantum generalization of this measure, an adjacency matrix of the quantum mutual information calculated on quantum states~\cite{NielsenChuangBook}, to well known quantum many-body models on regular 1D lattices, and uncover emergent quantum complexity which clearly identifies quantum critical points (QCPs)~\cite{carr2010k,sachdev2011quantum}.  Quantum mutual information bounds two-point correlations from above~\cite{Wolf2008}, measurable in a precise and tunable fashion in e.g. atom interferometry in 1D Bose gases~\cite{Langen2015}, among many other quantum simulator architectures. Using matrix product state (MPS) computational methods~\cite{openMPS,schollwoeck2011}, we demonstrate rapid finite size-scaling for both transverse Ising and Bose-Hubbard
models, including $Z_2$, mean field, and BKT quantum phase transitions.

As we move toward more and more complex quantum systems in materials design and quantum simulators, involving a hierarchy of scales, diverse interacting components, and a structured environment, we expect to observe long-lived dynamical features, fat-tailed distributions, and other key identifiers of complexity~\cite{complexityNAS2009,loganHillberryThesis2016,davidVargasThesis2016}.  Such systems include quantum simulator technologies based on ultracold atoms and molecules~\cite{carr2009b}, trapped ions~\cite{blatt2012}, and Rydberg gases~\cite{Loew2012}, as well as superconducting Josephson-junction based nanoelectromechanical systems in which different quantum subsystems form compound quantum machines with both electrical and mechanical components~\cite{Teufel2011}.  A key area in which we have taken a first step beyond phase diagrams and ground state properties is non-equilibrium quantum dynamics, where critical exponents and renormalization group theory are only weakly applicable at best, e.g. in the Kibble-Zurek mechanism, and are hard to find any use for at all in far-from-equilibrium regimes.  However, at the most basic level we can first ask, are quantum systems inherently complex?  Must we impose complexity on quantum systems to obtain it~\cite{Bianconi2001,Kimble2008,Cuquet2009,Perseguers2010,Bianconi2012a,Bianconi2012b}, or is there a regime in which complexity naturally emerges, even in ground states of regular lattice models?  In this Letter we show that emergent complexity can be well quantified in the simplest of 1D lattice quantum simulator models in terms of complexity measures around QCPs in direct analogy to similar measurements on the brain; moreover we establish a much-needed new set of tools for quantifying the complexity of far-from-equilibrium quantum dynamics.


\begin{figure}
\begin{center}
     \begin{overpic}[width=1.0 \columnwidth,unit=1mm]{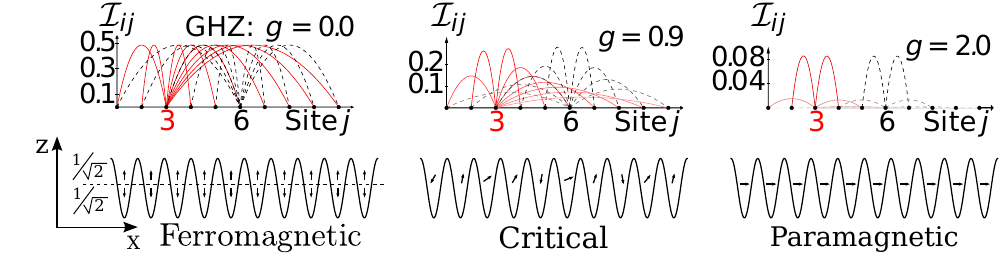}
        \put(0, 14){(a)}
        \put(0,1){(b)}
      \end{overpic}
\end{center}
\caption{\label{fig:sketch} \emph{Sketch of mutual information complex network.}  A chain of $L$ quantum bits for the transverse Ising model, the ``fruit-fly'' of quantum many body physics.  (a) Links originating from site 3 and site 6 for the mutual information complex network $\mathcal{I}_{ij}$, corresponding to phases and critical point in (b).  In this weighted network, the height of the links in our sketch denotes their relative strength; note descending vertical axes from left to right.  The entire complex network is far too dense to depict, so we show just two representative sites.  (b) Sketch of ferromagnetic phase, critical point, and paramagnetic phase.  The sinusoidal potential corresponds to an optical lattice for ultracold atoms or molecules.  In the ferromagnetic limit the ground state is the $Z_2$ symmetric superposition between all spin up and all spin down, indicated by two rows of arrows.}
\end{figure}

Quantum phase transitions are often characterized by quantum averages over physical observables such as two-point correlators.  For example, the transverse Ising model (TIM) consists of a chain of qubits with nearest neighbor $z$-$z$ coupling $J$ and a transverse field $g$. For large $g$ spins are not correlated in the $z$ direction, while for small $g$ the spins tend to align or anti-align, depending on the sign of $J$.  The quantum phase transition between large $g$ (paramagnetic) and small $g$ (ferro/antiferro-magnetic) at the QCP $g_c=1$ is evidenced by a change in the long range behavior of the two-point correlator $g^{(2)}_{ij}=\langle \hat{\sigma}_i^z \hat{\sigma}_j^z \rangle - \langle \hat{\sigma}_i^z \rangle \langle \hat{\sigma}_j^z \rangle $, where $i,j$ are sites on a lattice and $\hat{\sigma}^z$ are measurements of spin in the $z$-direction; alternate measures include the von Neumann entropy and concurrence~\cite{Amico2008}.  The mutual information $\mathcal{I}_{ij}$ is bounded from below by $g^{(2)}_{ij}$, and indeed by any possible two-point correlator in the model~\cite{Wolf2008}.  In general for quantum simulator technologies we obtain Hamiltonians for which we do not know \emph{a priori} what the right correlator is or indeed if there is a quantum phase transition at all.  Thus mutual information provides a much more general tool to identify such quantum phase transitions than any particular physical correlator.

To establish the usefulness of mutual information complex networks, we consider both the TIM and the Bose-Hubbard model (BHM).  The BHM balances particle tunneling $J$ against on-site particle interaction $U$, with the filling factor controlled by the chemical potential $\mu$; thus it has a richer phase diagram than the TIM, and exhibits both mean field transitions from Mott insulators to a superfluid phase as well as Berzinskii-Kosterlitz-Thouless (BKT) crossovers at commensurate filling.  We emphasize that both these models are studied heavily in quantum simulators experimentally and theoretically~\cite{lewensteinM2007,carr2009b,blatt2012,Britton2012,yanB2013}.

\textit{Quantum Many-body Hamiltonians and Mutual Information} --
The 1D transverse Ising model (TIM) takes the form
\begin{equation}\label{eqn:ising}
\hat{H}_{I} = -J\sum_{i=1}^{L-1}\hat{\sigma}_{i}^{z}\hat{\sigma}_{i+1}^{z} - J g
\sum_{
i=1}^{L} \hat{\sigma}_{i}^{x}\,,
\end{equation}
where $\commutator{\pauli{j}{\alpha}}{\pauli{k}{\beta}}=2i\delta_{jk}\epsilon_{\alpha\beta\gamma}\pauli{k}{\gamma}$.  The 1D Bose-Hubbard model (BHM) takes the form
\begin{equation}\label{eqn:boseHubbard}
\Hamiltonian{B}= -J\sum_{i=1}^{L-1}(\hat{b}^{\dagger}_{i} \hat{b}_{i+1}+
\text{h.c.}) +\frac{U}{2}\sum_{i=1}^{L}
\hat{n}_i(\hat{n}_i-\1) -\mu\sum_{i=1}^{L} \hat{n}_i\,,
\end{equation}
where $\commutator{\hat{b}_i}{\hat{b}^{\dagger}_j}=\delta_{ij}$ are bosonic annihilation and creation operators and $\hat{n}_i=\hat{b}^{\dagger}_i\hat{b}_i$.  Both the TIM and BHM are standard workhorses of quantum many-body lattice physics~\cite{sachdev2011quantum}.
The quantum mutual information $\mathcal{I}_{ij} \equiv \frac{1}{2}\left(S_i +S_j-S_{ij}\right)$, with $\mathcal{I}_{ii}\equiv0$, is constructed from the one and two point von Neumann entropies $S_{i}=-\mathrm{Tr}\left(\hat{\rho}_{i}\log_d{\hat{\rho}_{i}}\right)$, $S_{ij}=-\mathrm{Tr}\left(\hat{\rho}_{ij}\log_d{\hat{\rho}_{ij}}\right)$,
with reduced density operators defined in terms of the partial trace as
$\hat{\rho}_{i}=\underset{k \neq i}{\mathrm{Tr}} \hat{\rho}$ and $\hat{\rho}_{ij}=\underset{k \neq i,j}{\mathrm{Tr}} \hat{\rho}$.
We take $d=2$ for the TIM (qubits) and $d=n_{\text{max}}+1$ for the BHM, since particles can pile up on site in the latter, with $n_{\text{max}}=4$, a truncation parameter.

\textit{Complex Network Measures} -- We use weighted generalizations of standard measures based on unweighted adjacency matrices~\cite{Newman2003}; a formal justification for and interpretation of this generalization procedure can be found in~\cite{Ahnert2007}. A primitive measure of a node's importance is the sum of the weights connecting it to other nodes in the network, $s_i\equiv \sum_{j=1}^L\mathcal{I}_{ij}\,,$
where, $s_i$ is called the strength of node $i$.  The disparity $Y_i$ of a node $i$ in a network with $L$ nodes is defined as a function over weighted connections to its neighbors,
\begin{equation}\label{eqn:disparity}
Y_i \equiv  \frac{1}{\left(s_i\right)^2}\sum_{j=1}^L
\left(\mathcal{I}_{ij}\right)^2 = \frac{\sum_{j=1}^L
\left(\mathcal{I}_{ij}\right)^2}{\left(\sum_{j=1}
^L \mathcal{I}_{ij}\right)^2}\, .
\end{equation}
Observe that if the mutual information between lattice sites adopts a constant value $\mathcal{I}_{ij}=a$, that $Y_i=a^2\left(L-1\right)/a^2\left(L-1\right)^2=1/\left(L-1\right)$, so that if a node has relatively uniform weights across its neighbors the disparity between nodes will be approximately $1/\left(L-1\right)$. On the other hand, if a particular $\mathcal{I}_{ij}$ takes on a dominant value $b$, then $Y_{i}\approx b^2/b^2 = 1$. The average disparity over all nodes in the network
is $Y\equiv \frac{1}{L}\sum_{i=1}^L Y_i\,.$
The clustering coefficient $C$ is three times the ratio of triangles (three mutually connected vertices) to connected triples in an unweighted network.  In our weighted network,
\begin{equation}\label{eqn:clustering}
C\equiv\frac{\text{Tr}(\mathbf{\mathcal{I}}^{\thinspace3})}{\sum_{j\neq i}^L \sum_{
i=1}^L [\mathbf{\mathcal{I}}^{\thinspace 2}]_{ij}}\,.
\end{equation}
%
The density $D$ is the average fraction of the $\binom{L}{2}$ links that are present in the network:
\begin{equation}\label{eqn:netDensity}
D\equiv \frac{1}{L(L-1)}\sum_{i=1}^L s_i\,.
\end{equation}
As the number of nodes in an unweighted network is allowed to approach infinity a network is said to be \textit{sparse} if $D\rightarrow 0$, and \textit{dense} if $D>0$ as the number of nodes in the network $L$ approaches infinity~\cite{Newman2003}. Finally, a technique for assessing the similarity between two nodes $i,j$ in a network is to compute the Pearson correlation coefficient between them,
\begin{equation}\label{eqn:pearsonCorr}
r_{ij}\equiv \frac{\sum_{k=1}^L
\left(\mathcal{I}_{ik}-\braket{\mathcal{I}_i}\right)\left(\mathcal{I}_{jk}
-\braket{\mathcal{I}_j}\right)}{\sqrt{\sum_{k=1}^L
\left(\mathcal{I}_{ik}-\braket{\mathcal{I}_i}\right)^2}\sqrt{\sum_{k=1}^L
\left(\mathcal{I}_{jk}-\braket{\mathcal{I}_j}\right)^2}}\,,
\end{equation}
with $\braket{\mathcal{I}_i}$ the average of $\mathcal{I}_{ij}$ over $j$. $r_{ij}$ is treating link weight as a random variable; the numerator of Eq.~\eqref{eqn:pearsonCorr} is the covariance of the weights of
node $i$ with the weights of node $j$, while the denominator is the standard deviation in the weights of node $i$ multiplied by the standard deviation in the weights of node $j$. To restrict our study we focus on the Pearson correlation coefficient between the middle two sites of the lattice.  These two nodes are spatially close to each other and far from boundaries, making them the most similar nodes in the network whose weights of connection are not strongly modified by boundary conditions; we thus choose $R\equiv r_{\frac{L}{2},{\frac{L}{2} + 1}}$.

\textit{Numerical Techniques} -- We obtain our data with our widely-used MPS open source code~\cite{openMPS}, a well-established algorithm~\cite{schollwoeck2011}.  The essence of the approach is data compression of a quantum many-body state onto a classical computer, using singular value decomposition.  The key convergence parameter is the bond dimension $\chi$, limiting the growth of spatial entanglement as defined by the truncated Schmidt number of the reduced density matrix~\cite{NielsenChuangBook}. We use bond dimensions of up to several hundred, which are adequate to establish the usefulness of our complex network measures to pin down QCPs, as is our aim (for extremely high accuracy calculations with bond dimensions in the thousands see~\cite{Ejima2011}). Our simulations are converged up to a variance tolerance of  $10^{-10}(10^{-8})$ in the TIM (BHM). Mesoscopic corrections have been explored for the BHM in detail in our previous work~\cite{Carr2010}.

\begin{figure}
 \begin{center}
   \begin{minipage}{0.5\linewidth}
      \begin{overpic}[width=1.0 \columnwidth,unit=1mm]{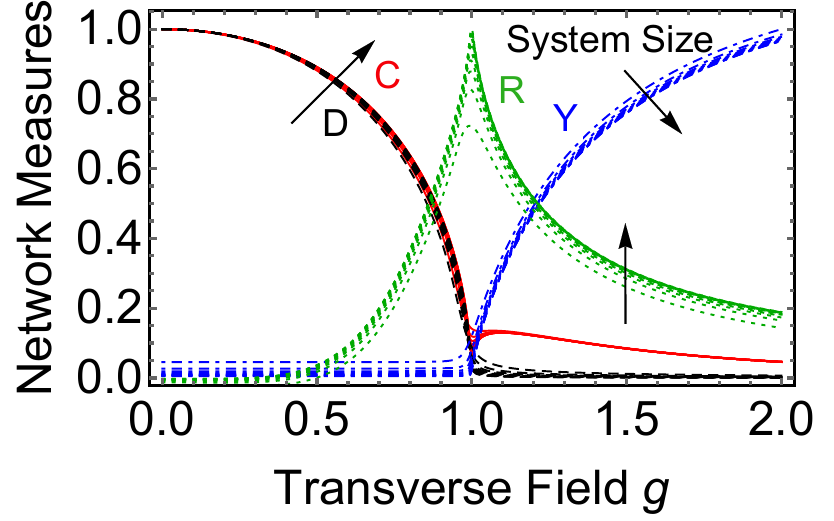}
        \put(2, 2){(a)}
      \end{overpic}
    \end{minipage}\hfill
    \begin{minipage}{0.5\linewidth}
      \begin{overpic}[width=1.0 \columnwidth,unit=1mm]{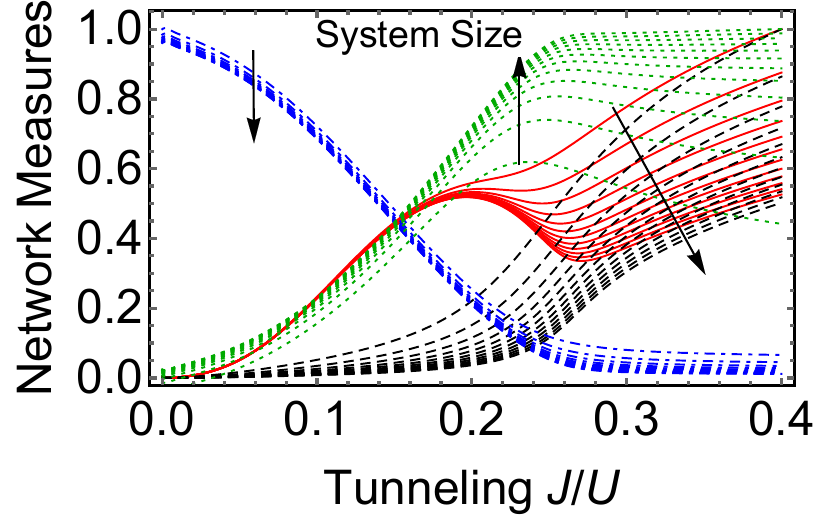}
        \put(2,2){(b)}
      \end{overpic}
    \end{minipage}\hfill
   \end{center}
\caption{\label{fig:splines} \emph{Complex network measures on the mutual information.} (a) Transverse quantum Ising model describing quantum spins (qubits).  The clustering coefficient $C$ and density $D$ serve as order parameters for the ferromagnetic phase.  The average disparity $Y$ identifies the short range correlations of the paramagnetic ground state. The Pearson correlation coefficient $R$ develops a cusp near the critical point $g_c=1$, identifying a structured nature to correlations near criticality. (b) Bose Hubbard model describing massive particles for commensurate lattice filling, with BKT crossover occurring in the limit $L \to \infty$ at a ratio of tunneling $J$ to interaction $U$ of $(J/U)_{\mathrm{BKT}}=0.305$; for smaller system sizes, the effective critical point~\cite{Carr2010} can be as small as $(J/U)_\mathrm{BKT}\simeq 0.2$.  The density and clustering coefficient grow as spatial correlations develop in the superfluid phase.  The average disparity is high in the Mott insulator phase where correlations are short-ranged.  Critical/crossover behavior is most evident in derivatives of these measures, see Fig.~\ref{fig:criticalPoint} and Table~\ref{tab:FitData}.  Note: all network measures have been self-normalized to unity for display on a single plot.}
\end{figure}

\textit{Emergence of Critical Points} -- Figure~\ref{fig:splines} shows a finite-size scaling study of complex network measures on the mutual information calculated with MPS code for these two models, for 1D lattices with ranges appropriate for experiments.  Although we studied twelve network measures, we selected the four most relevant for brevity: density of links $D$, clustering coefficient $C$, average disparity $Y$, and Pearson correlation between middle lattice sites, $R$, we also include other measures such as bond entropy $S_B$, negativity $\mathcal{N}$, correlation length $\xi$, and condensate depletion $\mathcal{D}$ for comparison purposes.  All four network measures are clearly useful to identify phase transitions in the TIM and highlight different physical aspects.  $D$ is high in the TIM ferromagnetic and BHM superfluid phases where the nodes in the lattice are strongly connected, as sketched in Fig.~\ref{fig:sketch}(a).  However, the quantum phase transition at the QCP is sharp at $L\to\infty$ for the TIM, where there is a $Z_2$ transition and $L\simeq 100$ suffices, whereas in the BHM we expect to observe a BKT crossover, which converges only for very large $L\simeq 1000$~\cite{Kuhner00}, and is most apparent in the first and second derivative of $D$.  The TIM paramagnetic and BHM Mott insulating phases are only sparsely connected.  $C$ follows a similar behavior except that for both the TIM and BHM it develops a local minimum near the QCP.  This reflects the fact that the average number of connected triples is temporarily growing faster the control parameter ($g$ for the TIM, $J/U$ for the BHM) for the average number of triangles.  Physically this could be because the length scale of correlations has become as long as one lattice spacing but not two, resulting in a period of rapid increase in mutual information between nearest neighbors relative to second nearest neighbors.  In strong contrast to $D$ and $C$, in the TIM ferromagnetic and BHM superfluid phases $Y$ asymptotically approaches $\frac{1}{L-1}$.  In the TIM paramagnetic and BHM Mott insulating phases, where correlations decay exponentially, $Y$ grows as spins become more tightly bound to their nearest neighbor relative to other qubits in the complex network.  Finally, $R$ has a completely different behavior, and clearly develops a cusp at the TIM QCP.  Qualitatively, $R$ is low in both the ferromagnetic and paramagnetic phases due to the relative homogeneity of correlations when $g \ll 1$ and when $g \gtrsim 2$. In contrast, near criticality the weights display an approximately linear relationship. 

\begin{figure}
 \begin{center}
   \begin{minipage}{0.5\linewidth}
      \begin{overpic}[width=1.0 \columnwidth,unit=1mm]{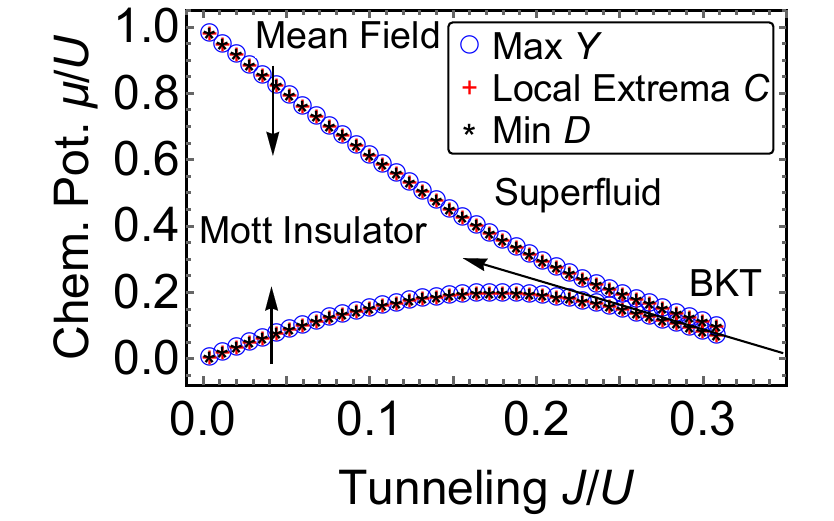}
        \put(2, 6){(a)}
      \end{overpic}
    \end{minipage}\hfill
    \begin{minipage}{0.5\linewidth}
      \begin{overpic}[width=1.0 \columnwidth,unit=1mm]{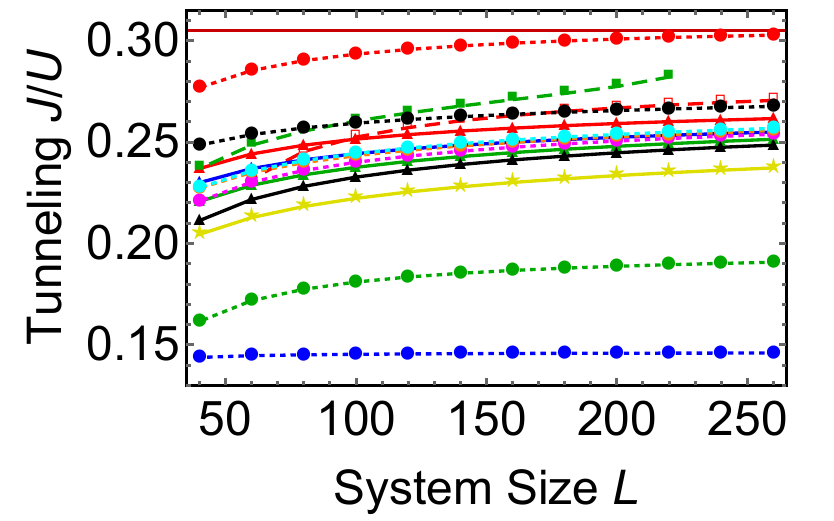}
        \put(2,6){(b)}
      \end{overpic}
    \end{minipage}\hfill
     \begin{minipage}{0.5\linewidth}
      \begin{overpic}[width=1.0 \columnwidth,unit=1mm]{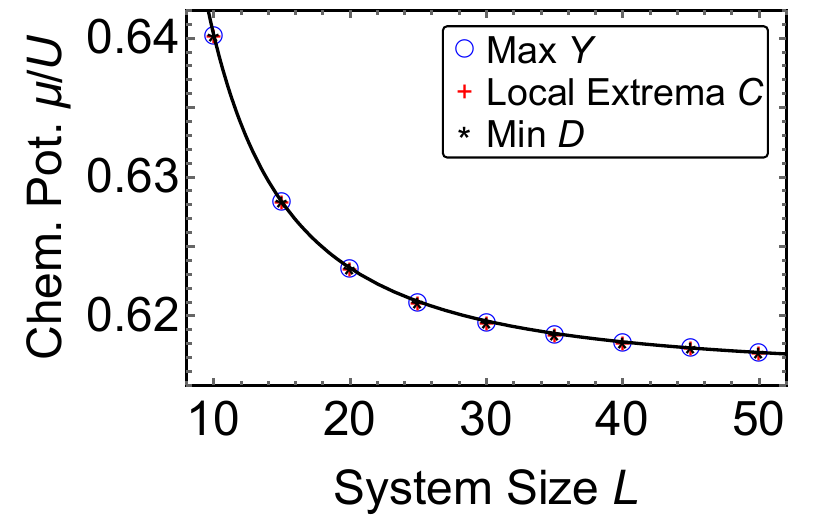}
        \put(2,6){(c)}
      \end{overpic}
    \end{minipage}\hfill
    \begin{minipage}{0.5\linewidth}
      \begin{overpic}[width=1.0 \columnwidth,unit=1mm]{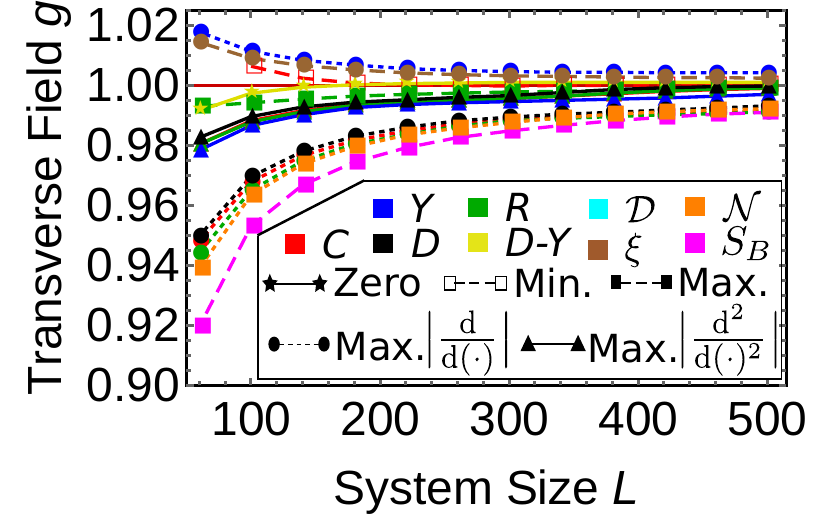}
        \put(2,6){(d)}
      \end{overpic}
    \end{minipage}\hfill
   \end{center}
\caption{\label{fig:criticalPoint} \emph{Finite-size scaling for the Bose-Hubbard model and transverse Ising model.}  (a)   BHM quantum phase diagram for fixed $L=40$ showing superfluid and Mott Insulating phases with mean field phase transition along the Mott lobe and a BKT transition at its tip.  (b) BHM BKT transition at unit filling. Scaling in $1/L$ places the critical point at $(J/U)_\mathrm{BKT} = 0.314$ (clustering $C$, dashed red), 0.281 (average disparity $Y$, solid blue) and 0.289 (density $D$, black dashed), respectively.  Compare to the best value to date~\cite{Ejima2011,Rigol2013} of 0.305, or the Luttinger liquid prediction of 0.328.  (c) Approaching the BHM mean field superfluid/Mott insulator transition for fixed $\left(J/U\right) = 0.1$.  Maximum disparity, central extrema in clustering, and minimum density scale towards the commensurate-incommensurate phase boundary and lie on top of each other.  (d) Scaling of multiple measures and their derivatives for the TIM, see also Table~\ref{tab:FitData}.}
\end{figure}

\textit{Finite-size scaling} --  Figure~\ref{fig:splines}(b) shows the BKT crossover transition for commensurate filling (average one particle per lattice site).  However, a mean field phase transition at non-commensurate filling also appears in the BHM.  As the Mott insulating phase is gapped (meaning the energy to create an excitation, even in the $L\to\infty$ limit, is non-zero), the usual way to find the boundaries of the Mott lobe (the region encompassing the Mott insulating phase) is to compute the energy required to add a particle or a hole to the insulator: the chemical potentials $\mu^p_c = E^p - E_0$ and $ -\mu^h_c = E^h - E_0$, respectively~\cite{Kuhner00}.  Then one uses finite-size scaling to extrapolate $\mu^p_c$ and $\mu^h_c$ in $L^{-1}$ to estimate the phase boundary.  Instead of working with chemical potentials, in Fig.~\ref{fig:criticalPoint}(a) we use $Y$ to obtain the first Mott lobe with both mean field and BKT crossover, shown here for $L=40$.  Figures~\ref{fig:criticalPoint}(b)-(c) show finite size scaling in $L$ towards the BKT crossover and the mean field phase transitions indicated in Fig.~\ref{fig:criticalPoint}(a).
Minimization of $D$ and tracking the central extrema of $C$, which goes from a global minimum to a local maximum, result in similar estimates to $Y$. The BKT transition has been estimated by many methods in the past, including from the correlator $\langle \bdag{i}\b{i+r}\rangle \sim r^{-K/2}$, taking advantage of the fact that at the QCP $K=1/2$~\cite{Kuhner00}, predicting $(J/U)_\mathrm{BKT} = 0.29 \pm 0.01$; more recent results estimate $(J/U)_\mathrm{BKT} = 0.305$~\cite{Ejima2011,Rigol2013}.
By fitting curves like those shown in Fig.~\ref{fig:criticalPoint}(b)-(c) (BHM) and Fig.~\ref{fig:criticalPoint}(d) (TIM), to power laws of the form $(J/U)_c\left(L\right)=(J/U)_c+A\,L^{-1/\nu'}$  (BHM) and $g_c\left(L\right)=g_c+A\,L^{-1/\nu}$ (TIM) we perform quantitative analysis of QCPs and provide errors due to the fitting procedure (see supplemental material for a detailed explanation) in Table~\ref{tab:FitData}.  In particular, examining this data we observe that by measuring the complex network structure present in the quantum mutual information, we can estimate the QCP of the TIM to within 0.01\% of its known value; that the Mott-insulator phase boundary can be reliably estimated by extremization of network quantities; and that the BKT transition at the tip of the Mott lobe, famously difficult to pin down without going to extremely large systems with 1000s of sites with high accuracy, is estimated to within 5.3\% using systems up to 260 sites via the max slope of $D$.

\begin{table}
\caption{\label{tab:FitData} \emph{Quantitative finite-size scaling analysis of quantum critical points.} Estimates for the critical point $g_c$ and $(J/U)_\mathrm{BKT}$ and scaling exponents $\nu,\nu'$ for the transverse Ising and Bose Hubbard models, respectively, based on \myst{three} complex network measures on the mutual information with standard quantum measures included for comparison.  We include first and second derivatives (F.D., S.D.) since bare measures are often insufficient, an effect well-known from one-point entanglement measures like the von Neumann entropy.  We also note two other features: the local minimum in the clustering coefficient $C$ (L.M.), and an intriguing point where normalized average disparity is equal to normalized density ($\tilde{Y} = \tilde{D}$).  Entries are left blank when the measure fails to identify the critical point.  Our complex network measures clearly perform as well or better than standard measures, particularly for the still improving estimates for the BHM BKT point~\cite{Ejima2011}.}
\begin{center}
\resizebox{\linewidth}{!} {
\begin{tabular}{cc|c|c|c|c}
Measure    & & $g_c$    & $\nu$    & $(J/U)_\text{BKT}$    & $\nu'$ \\ \hline
Density $D$ &F.D.   & $0.998\pm0.005$    & $0.962\pm0.245$    & $0.289\pm0.067$   & $2.980\pm6.642$    \\
Density $D$ &S.D.   & $1.005\pm0.011$    & $1.549\pm1.489$    & $0.287\pm0.055$   & $2.706\pm2.815$    \\
Disparity $Y$ &F.D.   & $1.003\pm0.004$    & $0.853\pm0.618$    &  &\\ 
Disparity $Y$ &S.D.   & $0.999\pm0.005$    & $0.972\pm0.597$    & $0.281\pm0.059$   & $2.809\pm4.529$    \\
Clustering $C$ &L.M.   & $1.000\pm0.001$    & $0.300\pm0.393$    & $0.281\pm0.012$   & $0.949\pm0.471$    \\
Clustering $C$ &F.D.   & $0.997\pm0.005$    & $0.954\pm0.237$    & $0.314\pm0.018$   & $1.538\pm1.343$    \\
Clustering $C$ &S.D.   & $1.003\pm0.008$    & $1.302\pm1.013$    & $0.281\pm0.041$   & $2.325\pm3.172$    \\
Pearson $R$ &F.D.   & $0.998\pm0.005$    & $0.988\pm0.232$    &    &     \\
Pearson $R$ &S.D.   & $1.005\pm0.012$    & $1.517\pm1.320$    & $0.300\pm0.111$   & $3.842\pm6.480$    \\
$\tilde{Y}=\tilde{D}$ &L.M.   & $1.001\pm0.002$    & $0.539\pm0.473$    & $0.299\pm0.148$   & $4.441\pm8.531$    \\
Bond Ent. $S_B$ &L.M.   & $1.000\pm0.005$    & $0.952\pm0.146$    &    &     \\
Bond Ent. $S_B$ &F.D.   &     &     & $0.283\pm0.047$    & $2.497\pm2.714$     \\
Negativity $\mathcal{N}$ &L.M.   & $1.000\pm0.005$    & $0.987\pm0.207$    &    &    \\
Negativity $\mathcal{N}$ &F.D.   &    &    & $0.289\pm0.069$    & $3.030\pm6.429$    \\
Corr. Len. $\xi$ &F.D.   & $1.001\pm0.005$   & $0.974\pm0.888$   &     &     \\
Depletion $\mathcal{D}$ &F.D.   &    &    & $0.286\pm0.054$    & $2.680\pm3.533$    \\

\end{tabular}
}
\end{center}
\end{table}

\textit{Conclusions} -- We have shown that quantum complexity already emerges in a clearly quantifiable way in quantum states near quantum phase transitions in regular 1D lattices. In direct analogy to complexity of EEG/fMRI measurements on the brain, our measures are built on taking the quantum mutual information as a weighted adjacency matrix, and reliably estimate quantum critical points for well-known quantum-many body models, in particular the transverse Ising and Bose-Hubbard models.  These models include three classes of phase transitions, $Z_2$, mean field superfluid/Mott insulator, and a BKT crossover; in each case we obtain rapidly converging accuracy for critical point values, a demonstrable improvement in finite-size scaling over all other known methods including e.g. high order perturbation theory. To be specific, the improvements come from the fact that the complex network measures do not require \textit{a priori} knowledge while retrieving important information about the quantum correlations. Our work sets the stage for application of a new set of quantum measures to quantify complexity of quantum systems where traditional correlation measures are at best weakly applicable.  In future work we will apply our new methods to far-from-equilibrium dynamics in such systems, for instance, quantum cellular automata~\cite{bleh2012,Arrighi2012,loganHillberryThesis2016,davidVargasThesis2016} and quantum degenerate ultracold molecules with a multiscale hierarchy of internal and external degrees of freedom.

\begin{acknowledgments}
The authors gratefully acknowledge valuable discussions with Justin Anderson, Arya Dhar, Wei Han, Logan Hillberry, David Larue, Kenji Maeda, Gavriil Shchedrin, and Michael Wall.  The authors also acknowledge the preliminary efforts of Adolfo Gomez. The calculations were carried out using the high performance computing resources provided by the Golden Energy Computing Organization at the Colorado School of Mines. This  material  is  based in  part  upon  work  supported  by  the  US National  Science Foundation  under  grant  numbers  PHY-1306638,  PHY-1207881, PHY-1520915, and OAC-1740130, and the US Air Force Office of Scientific Research grant number FA9550-14-1-0287, as well as a Colorado School of Mines graduate fellowship.  This work was also performed in part at the Aspen Center for Physics, which is supported by National Science Foundation grant PHY-1607611.
\end{acknowledgments}

\clearpage

\appendix
\section*{Supplementary Material}
In this supplementary material, we put the network measures in context
with other methods. In the first paragraph, we discuss common measures in
DMRG and tensor network methods to provide the reader with a comparison.
Then, we dedicate one paragraph to the computational cost of
the mutual information matrix in MPS methods. Finally, we provide the
detailed discussion of the error in the finite size scaling.

\begin{figure}[h!]
  \begin{center}
    \begin{minipage}{0.47\linewidth}
      \begin{overpic}[width=1.0 \columnwidth,unit=1mm]{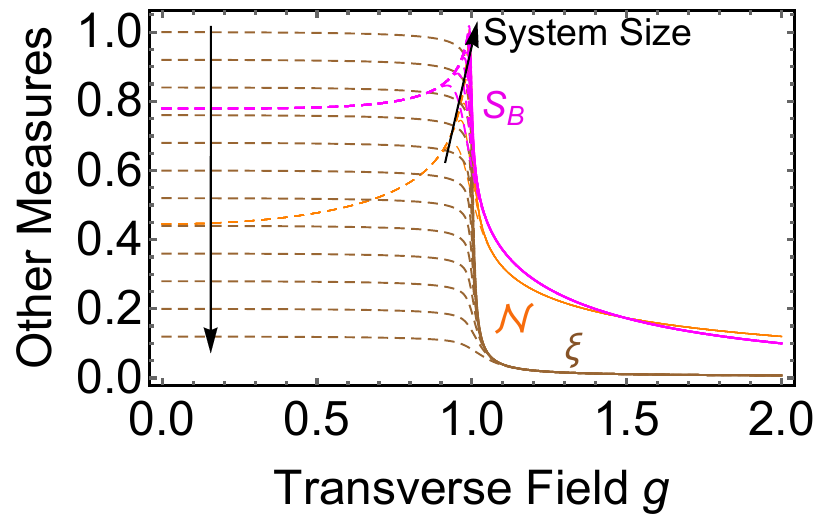}
        \put(0,0){(a)}
      \end{overpic}
    \end{minipage}\hfill
    \begin{minipage}{0.47\linewidth}
      \begin{overpic}[width=1.0 \columnwidth,unit=1mm]{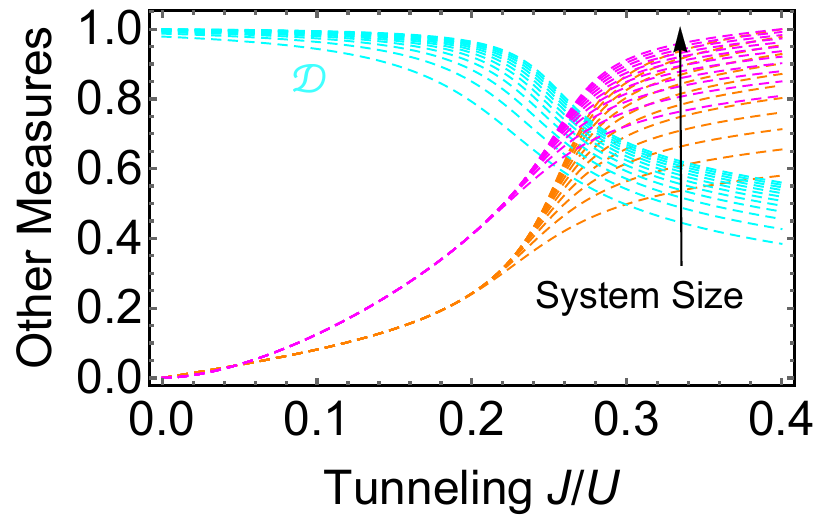}
        \put(0,0){(b)}
      \end{overpic}
    \end{minipage}

    \vspace{0.1cm}



    \begin{minipage}{0.47\linewidth}
      \begin{overpic}[width=1.0 \columnwidth,unit=1mm]{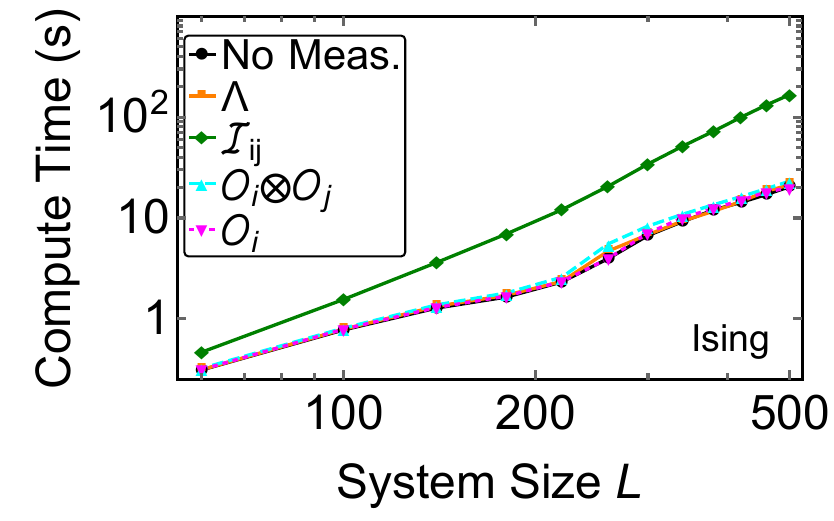}
        \put(0,0){(c)}
      \end{overpic}
    \end{minipage}\hfill
    \begin{minipage}{0.47\linewidth}
      \begin{overpic}[width=1.0 \columnwidth,unit=1mm]{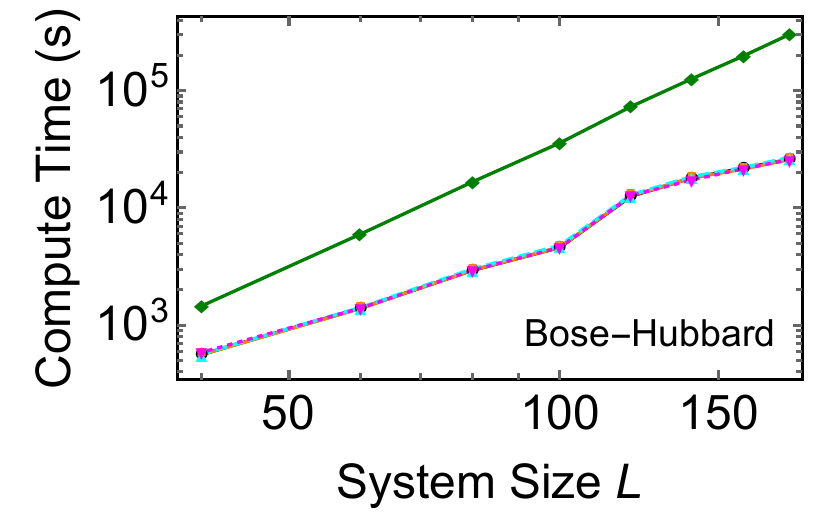}
        \put(0,0){(d)}
      \end{overpic}
    \end{minipage}

    \vspace{0.1cm}

    \begin{minipage}{0.47\linewidth}
      \begin{overpic}[width=1.0 \columnwidth,unit=1mm]{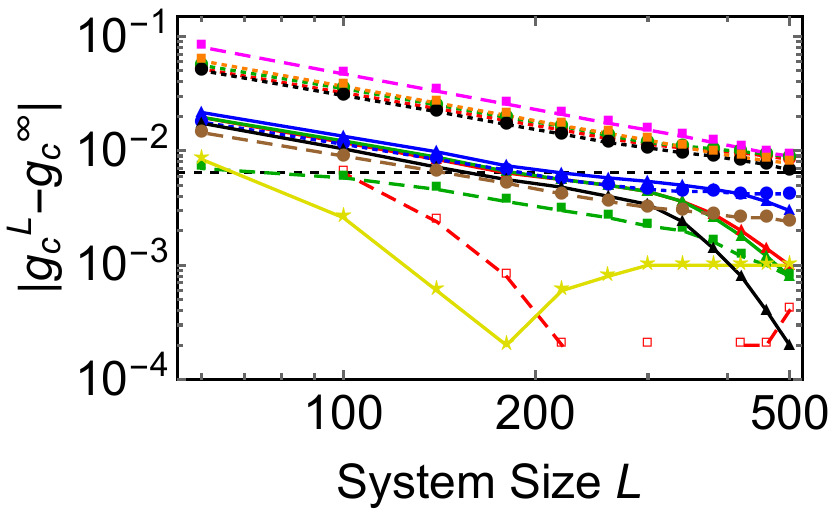}
        \put(0,0){(e)}
      \end{overpic}
    \end{minipage}\hfill
    \begin{minipage}{0.47\linewidth}
      \begin{overpic}[width=1.0 \columnwidth,unit=1mm]{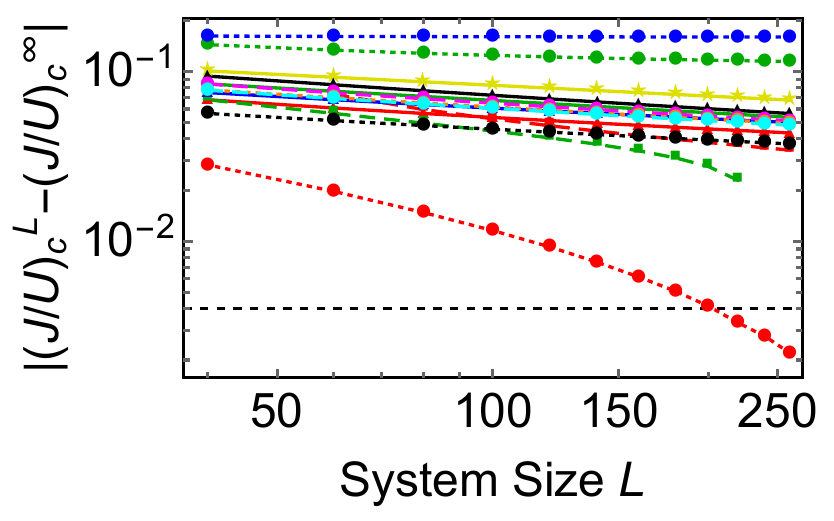}
        \put(0,0){(f)}
      \end{overpic}
    \end{minipage}
  \end{center}
  \caption{\emph{Supplemental plots.}
    (a)~Other measures, i.e., the bond entropy $S_{B}$,
    negativity $\mathcal{N}$, and correlation length $\xi$, for the
    quantum Ising model for different system sizes $L$ as a function
    of the external field $g$. (b)~$S_{B}$, $\mathcal{N}$, and
    depletion $\mathcal{D}$ for the Bose-Hubbard model
    at unit filling as a function of the tunneling strength $J / U$.
    (b) and (c) compare the other measures to the quantum network
    measures in Fig.~2 of the main body.
    (c)~Scaling of computation time for the ground including one
    selected measure: only ground state, eigenvalues $\Lambda$ of
    reduced density matrix for bipartition, mutual information matrix
    $\mathcal{I}_{ij}$, two-point correlators $O_i \otimes O_j$, and
    local observables $O_{i}$. (d)~Same as (c), but for the Bose-Hubbard
    model. (e)~log-log-plot of the extremas' and turning points' distance
    to the extrapolated $g_{c}^{\infty}$ of the variety of measures. The
    power-law behavior is clearly visible. The dashed line indicates the
    precision of the underlying grid. Peak of the yellow curve indicates
    that the measures is crossing $g_{c}^{\infty}$. Legends from Fig.~3(d)
    of the main body apply. (f)~Same as (e), but for the Bose-Hubbard
    model.
    }                                                              \label{fig:suppl}
\end{figure}

\emph{Common measures -} DMRG and MPS methods had means to find
the critical point of the ground state before quantum network measures.
To give a comparison, we have introduced such quantities in the
finite size scaling in Fig.~3 of the main body: we consider the bond
entropy $S_{B}$ and the negativity $\mathcal{N}$ as entanglement measures.
In the Ising model, we measure the correlation length $\xi$. The critical
point of the Bose-Hubbard model is evaluated with the depletion
$\mathcal{D}$. In analogy with Fig.~2 of our Letter, we plot these measures
for different system sizes in Fig.~\ref{fig:suppl}(a) and (b). $S_{B}$
is defined over a bipartition of the state; we choose sites $1, \ldots, L / 2$
and $L / 2 + 1, \ldots, L$. The eigenvalues of the reduced density matrix of
the partitions lead to $S_{B} = - \mathrm{Tr} \left( \hat{\rho}_{1, \ldots, L/2}
\log \left( \hat{\rho}_{1, \ldots, L/2} \right) \right)$. $\mathcal{N}$
is using the eigenvalues of the reduced density matrix $\Lambda_{\eta}$ for
the same bipartition $\hat{\rho}_{1, \ldots, L/2}$. For a pure state, we
obtain $\mathcal{N} = 1 / 2 \cdot \left(\left( \sum_{\eta, \eta'}
\sqrt{\Lambda_{\eta} \Lambda_{\eta'}} \right) - 1 \right)$. In the Ising model,
$\xi$ is calculated via the correlations $\langle \hat{\sigma}_{i}^{z}
\hat{\sigma}_{j}^{z} \rangle$ as
\begin{eqnarray}
  \xi =
  \sqrt{\frac{\sum_{l=1}^{L-1} l^2 \bar{c}_{l}}{\sum_{l=1}^{L-1} \bar{c}_{l}}}
  \, , \, \textrm{with } \bar{c}_{l} = \frac{1}{L-l} \left\|
          \sum_{j=1}^{L-l} \langle \sigma_{j}^{z} \sigma_{j+l}^{z} \rangle \right\| \, .
\end{eqnarray}
$\mathcal{D}$ in the Bose-Hubbard model is based on the single particle
density matrix (SPDM), i.e., the correlation matrix $\langle b_i^{\dagger}
b_j \rangle$. We calculate the eigenvalues $\lambda_{\eta}$ of the SPDM,
and define $\mathcal{D} = 1 - \max_{\eta} \lambda_{\eta} / \sum_{\eta}
\lambda_{\eta}$.

\emph{Computational scaling -} We present a short analysis of the
computational scaling of the network measures. We restrict ourselves
to the comparison within DMRG/MPS methods as a cross-method comparison
to, e.g., high-order perturbation theory, is difficult. The trend for
the quantum Ising model and the Bose-Hubbard model is that bond entropy
or local measures are faster than correlation measures based on
observables. The calculation of the two-site reduced density matrices
$\hat{\rho}_{ij}$ takes longer than the two previous measures, where
$\hat{\rho}_{ij}$ is necessary for the network measures.
Figure~\ref{fig:suppl}(c) and (d) support this statement. The quantum
Ising model is averaged over ten runs for each measurement. We list
the time for one run in the case of the Bose-Hubbard model. The
simulations are executed on a \emph{2x(Intel Xeon E5-2680 Dodeca-core)
24 Cores 2.50GHz} node. The
difference between the correlation measures and the two-sited reduced
density matrices is about one order of magnitude, or a factor of $10$
for the Ising model.
In the following, we present two arguments why we consider network
measures based on mutual information to be valuable. In the
example of the Ising model, we can use the $z$-$z$-correlations to
extract the quantum critical point. Indeed, this is faster than
calculating the network measures. For more complicated spin models,
it might be necessary to calculate all correlations and
cross-correlations summing up to nine or more correlation measurements.
Thus, calculating the full correlations requires about the same order
of magnitude of computation time. This argument becomes more
important moving from qubits to qudits, i.e., $d$-level systems.
For example, ten levels or more are necessary for many systems
in ultracold molecules \cite{Maeda2015}. The number of possible correlations and
cross-correlations inevitably grows, and a single number
extracted from network measures becomes more valuable. As a rule of
thumb, the number of cross-correlations will scale with $d^2$ in a
system of qudits. The additional computational effort of calculating
a reduced two-site density matrix over a single two-site correlator
is also $d^2$. Therefore, we consider network measures to be
advantageous for any system where many correlators have to be
calculated or when it is not clear which correlators will be useful
from the beginning.

\emph{Error analysis -} In the body of this Letter, Table~1
presents errors for the finite size scaling of each measure. We
give a detailed analysis of the error in this paragraph. As a
reminder of the data used for the finite size scaling, we present
the data used in a log-log-plot in Fig.~\ref{fig:suppl}(e) and
(f). These plots show the power-law behavior converging towards
the critical point for large system sizes $L$. The error given
in our results considers two sources of error. (i) the nonlinear
regression method used for the fit \cite{mathematicafit} has an
uncertainty inherent to the method. (ii) The extrema or turning points of the
quantities are based on an underlying grid of the external field
$g$ of the quantum Ising model, and the tunneling $J / U$ in the
Bose-Hubbard model, respectively. The discretization of these
parameters naturally leads to an error included as uncertainty
in the fitting procedure. These two sources are included in the
nonlinear regression. In addition,
we point out that each ground state and its measures have an
error due to the bond dimension bounding the maximally possible
entanglement \cite{osmpspaper2017}. This error should have the
same trend for neighboring values of $g$ and $J / U$ and is not
treated for this reason. (iv) The cut-off in the maximal filling
$n_{\max}$ in the Bose-Hubbard has been mentioned in the body of
the PRL as an additional source, where $\langle n_{\max} \rangle$
is negligible at an order of $10^{-3}$.


\begin{thebibliography}{10}

\bibitem{Newman2003}
M.~E.~J. Newman.
\newblock {The Structure and Function of Complex Networks}.
\newblock {\em SIAM Review}, 45:167--256, 2003.

\bibitem{Bianconi2001}
Ginestra Bianconi and Albert-L\'aszl\'o Barab\'asi.
\newblock {Bose-Einstein Condensation in Complex Networks}.
\newblock {\em Phys. Rev. Lett.}, 86:5632--5635, Jun 2001.

\bibitem{Kimble2008}
H.~J. Kimble.
\newblock The quantum internet.
\newblock {\em Nature}, 453:1023--1030, 2008.

\bibitem{Cuquet2009}
Mart\'i Cuquet and John Calsamiglia.
\newblock Entanglement percolation in quantum complex networks.
\newblock {\em Phys. Rev. Lett.}, 103:240503, Dec 2009.

\bibitem{Perseguers2010}
S.~Perseguers, M.~Lewenstein, A.~Ac\'{\i}n, and J.~I. Cirac.
\newblock {Quantum complex networks}.
\newblock {\em Nature Phys.}, 6:539--543, 2010.

\bibitem{Bianconi2012a}
Ginestra Bianconi.
\newblock Superconductor-insulator transition on annealed complex networks.
\newblock {\em Phys. Rev. E}, 85:061113, Jun 2012.

\bibitem{Bianconi2012b}
Arda Halu, Luca Ferretti, Alessandro Vezzani, and Ginestra Bianconi.
\newblock Phase diagram of the bose-hubbard model on complex networks.
\newblock {\em EPL (Europhysics Letters)}, 99(1):18001, 2012.

\bibitem{Bullmore2009}
Ed~Bullmore and Olaf Sporns.
\newblock Complex brain networks: graph theoretical analysis of structural and
  functional systems.
\newblock {\em Neuroscience}, 10:186, 2009.

\bibitem{NielsenChuangBook}
Michael~A Nielsen and Isaac~L Chuang.
\newblock {\em {Quantum Computation and Quantum Information: 10th Anniversary
  Edition}}.
\newblock Cambridge University Press, New York, NY, 2011.

\bibitem{carr2010k}
L.~D. Carr, editor.
\newblock {\em Understanding Quantum Phase Transitions}.
\newblock Taylor \& Francis, Boca Raton, FL, 2010.

\bibitem{sachdev2011quantum}
S.~Sachdev.
\newblock {\em Quantum Phase Transitions}.
\newblock Cambridge University Press, New York, 2011.

\bibitem{Wolf2008}
M.~M. Wolf, F.~Verstraete, M.~B. Hastings, and J.~I. Cirac.
\newblock Area laws in quantum systems: {M}utual information and correlations.
\newblock {\em Phys. Rev. Lett.}, 100:070502, 2008.

\bibitem{Langen2015}
T.~{Langen}, S.~{Erne}, R.~{Geiger}, B.~{Rauer}, T.~{Schweigler}, M.~{Kuhnert},
  W.~{Rohringer}, I.~E. {Mazets}, T.~{Gasenzer}, and J.~{Schmiedmayer}.
\newblock {Experimental observation of a generalized Gibbs ensemble}.
\newblock {\em Science}, 348:207--211, April 2015.

\bibitem{openMPS}
Matrix product state open source code.
  http://sourceforge.net/projects/openmps/.

\bibitem{schollwoeck2011}
Ulrich Schollwoeck.
\newblock The density-matrix renormalization group in the age of matrix product
  states.
\newblock {\em Ann. Phys.}, 326:96--192, 2011.

\bibitem{complexityNAS2009}
{\em The National Academies Keck Futures Initiative: Complex Systems: Task
  Group Summaries}.
\newblock The National Academies Press, Washington, D.C., 2009.

\bibitem{loganHillberryThesis2016}
Logan~E. Hillberry.
\newblock Entanglement and complexity in quantum elementary cellular automata.
\newblock Master's thesis, Colorado School of Mines, 2016.
\newblock
  http://inside.mines.edu/$\sim$lcarr/theses/hillberry\_thesis\_2016.pdf.

\bibitem{davidVargasThesis2016}
David~L. Vargas.
\newblock Quantum complexity: Quantum mutual information, complex networks, and
  emergent phenomena in quantum cellular automata.
\newblock Master's thesis, Colorado School of Mines, 2016.
\newblock http://inside.mines.edu/$\sim$lcarr/theses/vargas\_thesis\_2016.pdf.

\bibitem{carr2009b}
L.~D. Carr, D.~Demille, R.~V. Krems, and Jun Ye.
\newblock Cold and ultracold molecules: Science, technology, and applications.
\newblock {\em New J. Phys.}, 11:055049, 2009.

\bibitem{blatt2012}
R.~Blatt and C.~F. Roos.
\newblock {Quantum simulations with trapped ions}.
\newblock {\em {Nature Phys.}}, {8}:{277--284}, {2012}.

\bibitem{Loew2012}
Robert Loew, Hendrik Weimer, Johannes Nipper, Jonathan~B. Balewski, Bjoern
  Butscher, Hans~Peter Buechler, and Tilman Pfau.
\newblock {An experimental and theoretical guide to strongly interacting
  Rydberg gases}.
\newblock {\em {J. Phys. B. - At. Mol. Opt. Phys.}}, {45}({11}), {JUN 14}
  {2012}.

\bibitem{Teufel2011}
J.~D. Teufel, T.~Donner, Dale Li, J.~W. Harlow, M.~S. Allman, K.~Cicak, A.~J.
  Sirois, J.~D. Whittaker, K.~W. Lehnert, and R.~W. Simmonds.
\newblock {Sideband cooling of micromechanical motion to the quantum ground
  state}.
\newblock {\em {Nature}}, {475}({7356}):{359--363}, {JUL 21} {2011}.

\bibitem{Amico2008}
Luigi Amico, Rosario Fazio, Andreas Osterloh, and Vlatko Vedral.
\newblock Entanglement in many-body systems.
\newblock {\em Rev. Mod. Phys.}, 80:517--576, May 2008.

\bibitem{lewensteinM2007}
A.~Lewenstein, M.~Sanpera, V.~Ahufinger, B.~Damski, A.~Sen~De, and U.~Sen.
\newblock Ultracold atomic gases in optical lattices: Mimicking condensed
  matter physics and beyond.
\newblock {\em Adv. Phys.}, 56:243--379, 2007.

\bibitem{Britton2012}
Joseph~W. Britton, Brian~C. Sawyer, Adam~C. Keith, C.~C.~Joseph Wang, James~K.
  Freericks, Hermann Uys, Michael~J. Biercuk, and John~J. Bollinger.
\newblock {Engineered two-dimensional Ising interactions in a trapped-ion
  quantum simulator with hundreds of spins}.
\newblock {\em {Nature}}, {484}({7395}):{489--492}, {APR 26} {2012}.

\bibitem{yanB2013}
B.~{Yan}, S.~A. {Moses}, B.~{Gadway}, J.~P. {Covey}, K.~R.~A. {Hazzard}, A.~M.
  {Rey}, D.~S. {Jin}, and J.~{Ye}.
\newblock {Realizing a lattice spin model with polar molecules}.
\newblock {\em Nature}, 492:396, May 2013.

\bibitem{Ahnert2007}
S.~E. Ahnert, D.~Garlaschelli, T.~M.~A. Fink, and G.~Caldarelli.
\newblock Ensemble approach to the analysis of weighted networks.
\newblock {\em Phys. Rev. E}, 76:016101, Jul 2007.

\bibitem{Ejima2011}
S.~{Ejima}, H.~{Fehske}, and F.~{Gebhard}.
\newblock {Dynamic properties of the one-dimensional Bose-Hubbard model}.
\newblock {\em Europhys. Lett.}, 93:30002, February 2011.

\bibitem{Carr2010}
L.~D. Carr, M.~L. Wall, D.~G. Schirmer, R.~C. Brown, J.~E. Williams, and
  Charles~W. Clark.
\newblock {Mesoscopic effects in quantum phases of ultracold quantum gases in
  optical lattices}.
\newblock {\em Phys. Rev. A}, 81(1):013613, January 2010.

\bibitem{Kuhner00}
Till~D. K\"uhner, Steven~R. White, and H.~Monien.
\newblock One-dimensional bose-hubbard model with nearest-neighbor interaction.
\newblock {\em Phys. Rev. B}, 61:12474--12489, May 2000.

\bibitem{Rigol2013}
M.~Rigol.
\newblock Scaling of the gap, fidelity susceptibility, and bloch oscillations
  across the superfluid-to-mott-insulator transition in the one-dimensional
  bose-hubbard model.
\newblock {\em Phys. Rev. A}, 87:043606, 2013.

\bibitem{bleh2012}
D.~Bleh, T.~Calarco, and S.~Montangero.
\newblock {Quantum Game of Life}.
\newblock {\em {Europhys. Lett.}}, {97}({2}), {JAN} {2012}.

\bibitem{Arrighi2012}
Pablo Arrighi and Jonathan Grattage.
\newblock {The quantum game of life}.
\newblock {\em {Phys. World}}, {25}({6}):{23--26}, {JUN} {2012}.

\bibitem{mathematicafit}
We use the \emph{mathematica} function "NonlinearModelFit".

\bibitem{osmpspaper2017}
D.~{Jaschke}, M.~L. {Wall}, and L.~D. {Carr}.
\newblock {Open source Matrix Product States: Opening ways to simulate
  entangled many-body quantum systems in one dimension}.
\newblock {\em ArXiv e-prints 1703.00387}, Mar 2017.

\bibitem{Maeda2015}
Kenji Maeda, Michael~L. Wall, and Lincoln~D. Carr.
\newblock {Hyperfine structure of the hydroxyl free radical (OH) in electric
  and magnetic fields}.
\newblock {\em New Journal of Physics}, 17(4):045014, 2015.

\end{thebibliography}

\end{document}